\pdfoutput=1
\documentclass{article}

\IfFileExists{refs.bib}{}{}
\makeatletter\def\input@path{{arxiv/}}\makeatother

\usepackage{arxiv}

\renewcommand{\shorttitle}{}

\usepackage{etoolbox}
\makeatletter
\patchcmd{\@maketitle}{\textsc{\undertitle}\\}{}{}%
  {\PackageError{main}{Could not remove the undertitle}{}}
\makeatother

\makeatletter
\renewcommand{\section}{\@startsection{section}{1}{\z@}%
  {-2.0ex}{1.5ex}{\large\bf\raggedright}}
\renewcommand{\subsection}{\@startsection{subsection}{2}{\z@}%
  {-1.8ex}{0.8ex}{\normalsize\bf\raggedright}}
\makeatother

\usepackage[utf8]{inputenc} 
\usepackage[T1]{fontenc}    
\usepackage{hyperref}       
\usepackage{url}            
\usepackage{booktabs}       
\usepackage{amsfonts}       
\usepackage{amsmath}
\usepackage{mathtools}
\usepackage{microtype}      
\usepackage{graphicx}
\graphicspath{{./}{arxiv/}}
\usepackage{float}
\usepackage[numbers,sort&compress]{natbib}
\usepackage[skip=7pt]{caption}
\usepackage{tikz}
\usetikzlibrary{arrows.meta,positioning,shapes.geometric,fit,backgrounds}

\newcommand{\DP}{\mathrm{DP}}
\newcommand{\GEQO}{\mathrm{GEQO}}
\newcommand{\MCTS}{\mathrm{MCTS}}
\newcommand{\PG}{\mathrm{PG}}
\newcommand{\Mcal}{\mathcal{M}}

\title{Probe-Based Routing for Join Optimizer Selection}

\author{
Vladimir Burlakov\\
Lomonosov Moscow State University\\
\texttt{burlakovvs@my.msu.ru}\\
\and
\textbf{Alena Rybakina}\\
Innopolis University\\
\and
\textbf{Sergey Kudashev}\\
Innopolis University\\
\and
\textbf{Ilgam Latypov}\\
Lomonosov Moscow State University\\
\and
\textbf{Konstantin Gilev}\\
Novosibirsk State University\\
\and
\textbf{Alexander Demin}\\
Ershov Institute of Informatics Systems\\
\and
\textbf{Denis Ponomaryov}\\
Ershov Institute of Informatics Systems\\
\and
\textbf{Yuriy Dorn}\\
Lomonosov Moscow State University\\
}

\hypersetup{pdftitle={Probe-Based Routing for Join Optimizer Selection}}

\begin{document}
\maketitle

\begin{abstract}
In optimizers that choose a join-order solver by a size rule, join ordering is an algorithm-selection problem, because dynamic programming (DP), genetic search (GEQO), and Monte Carlo tree search (MCTS) each win on different queries. PostgreSQL nevertheless chooses among them by relation count alone, although queries with the same relation count can favor different solvers. This paper introduces two landscape features that a probe computes from random join orders in microseconds per query. The tail spacing $\alpha$ describes how the costs of the best sampled orders are spaced, and the crossover surprise $\theta$ describes how far the cost of a crossover child deviates from the mean cost of its parents. The router clusters training queries by these features and the relation count, assigns a solver to every cluster that this solver wins in total end-to-end time by a margin, and leaves all other queries to PostgreSQL's default rule. On JOB and IMDb-CEB, the router lowers mean end-to-end time relative to the default rule, and on IMDb-CEB the probe features nearly double the gain obtained from the relation count alone, so they carry information about the queries that the relation count lacks.
\end{abstract}

\section{Introduction}
\label{sec:intro}

A query optimizer with several join-order solvers must decide which of them
receives a query before join enumeration begins. PostgreSQL makes this
decision with its GEQO threshold. By default, queries with fewer than 12
relations go to dynamic programming (DP) and larger ones to the genetic
optimizer GEQO~\cite{postgresql2025geqo}. The rule is free to evaluate, but it
sees nothing of a query except its relation count.

The choice matters for end-to-end time, which includes both planning and
execution. On JOB, JOB-Complex, and IMDb-CEB, each of DP, GEQO, and the Monte Carlo tree
search (MCTS) solver of~\cite{burlakov2026practical} is the fastest on roughly
a quarter to a half of the queries (Table~\ref{tab:baselines}). Even
PostgreSQL's two native solvers split JOB almost evenly, with DP ahead of GEQO
on 60 of 113 queries. Because one slow choice on an expensive query can
outweigh many correct choices on cheap ones, a router has to be judged by
total end-to-end time rather than by the share of queries it routes
correctly.

Existing systems make this choice in one of two ways. Deployed optimizers
apply size rules, as PostgreSQL does with its GEQO threshold, Spark SQL with a
relation-count limit on DP join reordering, and Umbra with its switch among
DPhyp, linearized DP, and greedy search~\cite{neumann2018adaptive}. Learned
selectors predict the choice from query features instead. AlphaJoin~2.0 learns
a per-query choice between a learned MCTS solver and PostgreSQL's
optimizer~\cite{zhang2020alphajoin}, and SATzilla-style portfolios predict the
runtime of each solver and run the fastest one~\cite{xu2008satzilla}.

Size rules cannot separate queries of equal size. JOB queries 28a and 33c both
join 14 relations, so PostgreSQL sends both to GEQO, yet GEQO is the slowest
solver on 28a and beats DP on 33c (Section~\ref{sec:probe}). Join-graph
features do not resolve such cases either, because they describe the shape of
the graph and not how plan costs are distributed over join orders, which is
what governs the behavior of randomized solvers such as GEQO and MCTS.

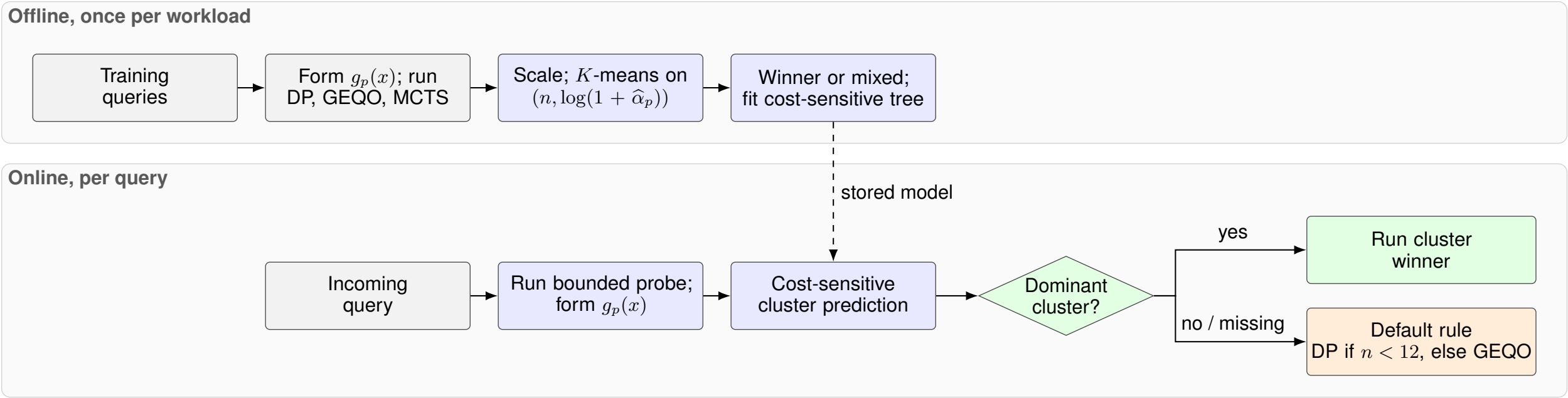
\begin{figure*}[t]
\centering
\resizebox{\textwidth}{!}{%
\begin{tikzpicture}[
  font=\sffamily\small,
  base/.style={draw=black!70, rounded corners=2pt, align=center,
               text width=32mm, minimum height=11mm, inner sep=2pt},
  data/.style={base, fill=gray!10},
  learned/.style={base, fill=blue!9},
  outcome/.style={base, text width=36mm, fill=green!11},
  fallback/.style={base, text width=36mm, fill=orange!15},
  decision/.style={draw=black!70, diamond, aspect=2.2, inner sep=1pt,
                   align=center, fill=green!11},
  arrow/.style={-{Latex[length=2mm]}, semithick},
  lane/.style={draw=gray!35, fill=gray!4, rounded corners=4pt},
  lanelabel/.style={font=\sffamily\small\bfseries, text=black!65,
                    anchor=north west, inner sep=3pt}]
  \node[data]    (train)   at (0,0)     {Training\\queries};
  \node[data]    (measure) at (3.8,0)   {Form $g_p(x)$; run\\DP, GEQO, MCTS};
  \node[learned] (cluster) at (7.6,0)   {Scale; $K$-means on\\$(n,\log(1+\widehat\alpha_p))$};
  \node[learned] (fit)     at (11.4,0)   {Winner or mixed;\\fit cost-sensitive tree};
  \node[data]     (query)   at (3.8,-3.4) {Incoming\\query};
  \node[learned]  (probe)   at (7.6,-3.4) {Run bounded probe;\\form $g_p(x)$};
  \node[learned]  (predict) at (11.4,-3.4) {Cost-sensitive\\cluster prediction};
  \node[decision] (dom)     at (15.2,-3.4) {Dominant\\cluster?};
  \node[outcome]  (win)     at (21.0,-2.65) {Run cluster\\winner};
  \node[fallback] (pg)      at (21.0,-4.15) {Default rule\\DP if $n<12$, else GEQO};
  \draw[arrow] (train) -- (measure);
  \draw[arrow] (measure) -- (cluster);
  \draw[arrow] (cluster) -- (fit);
  \draw[arrow] (query) -- (probe);
  \draw[arrow] (probe) -- (predict);
  \draw[arrow] (predict) -- (dom);
  \draw[arrow] (dom.east) -- ++(0.35,0) |- node[pos=0.72, above, font=\sffamily\footnotesize]{yes} (win.west);
  \draw[arrow] (dom.east) -- ++(0.35,0) |- node[pos=0.72, above, font=\sffamily\footnotesize]{no / missing} (pg.west);
  \draw[arrow, dashed] (fit.south) -- node[right, font=\sffamily\footnotesize]{stored model} (predict.north);
  \coordinate (offtop) at ([yshift=5mm]fit.north);
  \coordinate (ontop) at ([yshift=5mm]win.north);
  \begin{scope}[on background layer]
    \node[lane, fit=(train)(fit)(win.east |- fit.north)(offtop),
          inner xsep=5mm, inner ysep=3.5mm] (offlane) {};
    \node[lane, fit=(train.west |- win.north)(pg)(query)(ontop),
          inner xsep=5mm, inner ysep=3.5mm] (onlane) {};
  \end{scope}
  \node[lanelabel] at (offlane.north west) {Offline, once per workload};
  \node[lanelabel] at (onlane.north west) {Online, per query};
\end{tikzpicture}}
\caption{The router. Offline, each workload receives its own scaling,
clusters, dominance labels, and tree. Online, a query in a dominant cluster
runs with the cluster winner, and every other query keeps PostgreSQL's
default rule.}
\label{fig:architecture}
\end{figure*}

The router proposed here uses two features of that distribution
(Figure~\ref{fig:architecture}). A probe scores random left-deep join
orders with an estimated cost and derives the tail spacing $\widehat\alpha_p$
and the crossover surprise $\widehat\theta_p$ from these scores in at most
0.06\,ms per query (Section~\ref{sec:probe}). Offline, the router clusters the
training queries of a workload by these features and the relation count. A
cluster that one solver wins in total end-to-end time by a margin, and that
spans enough query templates, is dominant and stores this solver, while all
other clusters are mixed. Online, a shallow tree maps an incoming query to a
cluster, and the query either runs with the solver of its dominant cluster or
keeps PostgreSQL's default rule. Queries of the second kind are planned
exactly as in stock PostgreSQL, so every difference from the default comes
from the solvers that dominant clusters store.

The evaluation replays recorded end-to-end times on 50 random 80/20 splits of
each workload, with one router configuration that is chosen on JOB and
reused unchanged on all three workloads. The router lowers mean end-to-end
time relative to the default rule by 12.3\% on JOB and 8.5\% on IMDb-CEB
(Table~\ref{tab:primary}), and on IMDb-CEB the probe features nearly double
the gain obtained from the relation count alone
(Table~\ref{tab:ceb-ablation-mcts}). On JOB-Complex, whose 29 queries are
hand-crafted hard cases, the router stays within 3\% of the default. Simple
policies on the same splits, such as always running MCTS or a learned
relation-count threshold between DP and MCTS, show how much of the gain comes
from the larger portfolio and how much from routing
(Table~\ref{tab:controls}).

The paper makes four contributions.

\begin{enumerate}
\item Two landscape features for solver selection, the tail spacing and the
crossover surprise, which a probe computes in at most 0.06\,ms per query.

\item A router that assigns a solver only to dominant clusters of these
features and leaves all other queries to PostgreSQL's default rule.

\item An evaluation on JOB, JOB-Complex, and IMDb-CEB in which the router and
simple baseline policies share the same splits, which separates the effect of
adding MCTS, of moving the relation-count threshold, and of choosing the
solver per query.

\item A feature ablation on IMDb-CEB with and without MCTS in the portfolio.
\end{enumerate}

\section{Related Work}
\label{sec:related}

\subsection{Join-Order Search and Learned Optimization}

System~R introduced cost-based DP for left-deep join orders, and Volcano and
Cascades organized optimizer search around extensible
rules~\cite{selinger1979access,graefe1993volcano,graefe1995cascades}. Later DP
algorithms exploit join-graph connectivity and hypergraph structure to
enumerate bushy plans~\cite{moerkotte2006analysis,moerkotte2008dphyp}, while
randomized search, iterative DP, query simplification, and adaptive
linearization give up completeness in order to bound planning effort on large
joins~\cite{ioannidis1990randomized,kossmann2000iterative,neumann2009query,neumann2018adaptive}.
All of these methods change how a single solver searches, whereas this paper
asks which solver should receive a query.

Learned optimizers widen the set of actions available to the optimizer.
ReJOIN, Neo, and Balsa learn to construct join orders or
plans~\cite{marcus2018rejoin,marcus2019neo,yang2022balsa}, whereas Bao,
FASTgres, AutoSteer, and Lero keep the native optimizer and learn hint
selection or plan
preference~\cite{marcus2021bao,woltmann2023fastgres,anneser2023autosteer,zhu2023lero}.
Kepler generates a plan set offline and uses uncertainty-aware prediction to
avoid regressions, and GenJoin predicts how to transform an input set of
subplan hints~\cite{doshi2023kepler,sulimov2025genjoin}. Because these systems
choose plans, preferences, or hints after the solver has been chosen, they can
run on top of whichever solver the router selects.

AlphaJoin uses learned MCTS for join ordering, and HybridQO uses MCTS inside a
pipeline that generates and selects candidate
plans~\cite{zhang2020alphajoin,yu2022hybridqo}. Its successor AlphaJoin~2.0
learns for each query whether its MCTS solver or PostgreSQL's optimizer should
plan the query~\cite{zhang2020alphajoin}. The router described here chooses among three solvers instead of two and
decides per cluster of queries rather than per query. It assigns a solver to a cluster only after the
dominance test of Section~\ref{sec:clusters} and leaves the queries of all
other clusters to PostgreSQL's choice.

\subsection{Offline Optimization and Robustness}

Offline learned optimization moves expensive exploration out of query
planning. QO-Advisor, AutoSteer, and Kepler execute alternative plans offline
and keep the plan changes they have
verified~\cite{zhang2022qoadvisor,anneser2023autosteer,doshi2023kepler}.
BayesQO searches plans by Bayesian optimization with execution as an offline
step, and LimeQO models the matrix of query and hint latencies as low rank in
order to choose which entries to measure~\cite{tao2025bayesqo,yi2025limeqo}.
Both target recurring analytic workloads, in which the offline work is spread
over many executions. The router also executes alternatives offline, but it
stores a rule for choosing a solver instead of verified plans, so every new
query is planned from scratch without a plan tested offline.

Cost and cardinality estimates do not reliably predict the quality of the plan
that an optimizer finally selects~\cite{leis2015good}, and Izenov et al. show
that Q-error is a weak indicator of plan suboptimality, proposing a classifier
based on L1 error instead~\cite{izenov2024l1error}. The evaluation in this
paper therefore relies on measured end-to-end times rather than on estimated
costs. ROME executes complementary plans in parallel and returns the first
result~\cite{wei2024rome}, whereas the router runs a single solver per query,
so a wrong choice costs its full end-to-end time.

\subsection{Algorithm Selection and Portfolios}
\label{sec:related-portfolios}

Per-instance algorithm selection maps instance features to a solver under an
application-specific cost~\cite{rice1976algorithm}, and algorithm portfolios
exploit solvers that win on different instances~\cite{gomes2001portfolios}.
SATzilla made feature cost, timeouts, and cost-sensitive selection explicit,
and later variants are covered by
surveys~\cite{xu2008satzilla,kotthoff2014algorithm}. Solver selection in a
database shares this asymmetry, in which one slow choice can account for a
large share of workload time even when most choices are correct.

Several portfolio methods contain parts of the router. ISAC clusters
instances and assigns a configuration to each cluster~\cite{kadioglu2010isac},
exploratory landscape analysis combines sampled instance features with
cost-sensitive learning~\cite{bischl2012landscape}, and cost-sensitive
hierarchical clustering forms regions with a common
solver~\cite{malitsky2013cshc}. Hydra constructs complementary solver
configurations, confidence-aware recourse falls back to a default solver, and
AutoFolio configures the selector
itself~\cite{xu2010hydra,ansotegui2018recourse,lindauer2015autofolio}.
Clustering, cost-sensitive assignment, and fallback are thus well established,
and what is new here is their use for join-order solver selection, with
features computed during planning, end-to-end time as the loss, a separate
model for each workload, and PostgreSQL's default rule as the fallback.

\subsection{Query Regions, Plan Reuse, and Adaptivity}

Parametric query optimization partitions the parameter space into regions in
which different plans are
best~\cite{ioannidis1992parametric,hulgeri2002parametric}. Parametric plan
caching clusters parameter bindings that can share a plan, and PLASTIC
clusters past queries and classifies new ones for plan
reuse~\cite{aluc2012caching,ghosh2002plastic}. RankPQO ranks a compact plan
cache, Plan Stitch combines subplans observed in earlier executions, and PLARQ
retrieves and ranks candidates from a precomputed
pool~\cite{mo2024rankpqo,dutt2019plan,lan2026plarq}. These methods select or
build a plan, usually for recurring instances of a parameterized template,
whereas the clusters of the router group queries that share a fastest solver.
The chosen solver still plans every query from scratch, and the router does
not need to know the template of a query online.

Adaptive query processing acts after optimization has begun. Eddies route
tuples continuously, progressive optimization triggers reoptimization at
checkpoints, and SkinnerDB learns join orders from execution
progress~\cite{avnur2000eddies,markl2004progressive,trummer2019skinner}. The
router, in contrast, makes a single decision before join enumeration and
leaves both the chosen solver and the executor unchanged.

\section{Router Design}
\label{sec:design}

\subsection{Solvers and Default Rule}

The router chooses among three solvers, $\Mcal=\{\DP,\GEQO,\MCTS\}$, which
trade planning effort against plan quality in different ways. DP enumerates the configured
plan space, so its planning effort grows quickly with the number of
relations~\cite{selinger1979access,moerkotte2008dphyp}, whereas GEQO bounds this
effort by randomized genetic
search~\cite{ioannidis1990randomized,steinbrunn1997heuristic} and MCTS
searches partial join orders with tree exploration and cost-guided
rollouts~\cite{burlakov2026practical,kocsis2006bandit}. For query $i$ and
solver $m$, the router optimizes the end-to-end time
\[T_{im}=T^{\mathrm{plan}}_{im}+T^{\mathrm{exec}}_{im},\]
which includes planning because a solver that finds a slightly faster plan
can spend more time in planning than it saves in execution.

Stock PostgreSQL chooses between DP and GEQO by relation count alone. With
$n(x)$ the number of base relations that enter join enumeration for query $x$,
GEQO enabled, and \texttt{geqo\_threshold} at its default of 12, it uses
\begin{equation}
\pi_{\PG}(x)=
\begin{cases}
\DP, & n(x)<12,\\
\GEQO, & n(x)\ge 12.
\end{cases}
\label{eq:pg}
\end{equation}
This default rule is both the baseline of the evaluation and the fallback of
the router.

\subsection{Probe Features}
\label{sec:probe}

Features for this decision have to be computed during planning, before any
solver runs, and have to describe how plan costs are distributed over join
orders rather than how the join graph looks. The probe obtains such features
by scoring a few random left-deep join orders, which yields the probe
features
\begin{equation}
g_p(x)=\bigl(n(x),\widehat\alpha_p(x),\widehat\theta_p(x)\bigr).
\label{eq:features}
\end{equation}
For a random permutation $\pi$ of the
relations, let $\widehat r_j$ be the optimizer's cardinality estimate of base
relation $j$ and $e_k(\pi)$ the number of join-graph edges among the first $k$
relations. With the selectivity fixed at $\bar s=0.1$, the score of the order
is
\begin{equation}
\begin{aligned}
\ell_k(\pi)
  &=\sum_{j=1}^{k}\log\max(\widehat r_{\pi_j},1)
    +e_k(\pi)\log\bar s,\\
f(\pi)
  &=\log\sum_{k=2}^{n}\exp\ell_k(\pi).
\end{aligned}
\label{eq:probe-score}
\end{equation}
The score $f$ is the classical $C_{\mathrm{out}}$ cost, the sum of estimated
intermediate result sizes along the order, computed in log space from base
cardinalities and a single fixed selectivity. Working in log space avoids
overflow on large joins and makes scores comparable across queries.

The tail spacing summarizes the low end of the sampled cost distribution,
where the competitive orders lie. The probe draws $R=\max(2n,8)$ random orders
within a time cap and sorts their scores as $f_{(1)}\le\cdots\le f_{(R)}$. With $k=\lfloor\sqrt R\rfloor$ and
$J=\{i\in\{2,\ldots,k+1\}\mid f_{(i)}>f_{(1)}\}$, the tail spacing is
\begin{equation}
\widehat\alpha_p=
\left[
\frac{1}{|J|}\sum_{i\in J}
\log\frac{f_{(k+1)}-f_{(1)}}{f_{(i)}-f_{(1)}}
\right]^{-1},
\label{eq:probe-alpha}
\end{equation}
a Hill-style statistic of the spacing among the lowest sampled scores,
measured from the sample minimum. Before scaling, the router replaces $\widehat\alpha_p$ by $\log(1+\widehat\alpha_p)$
to reduce its range, which yields the model vector
\begin{equation}
z_p(x)=\bigl(n(x),\log(1+\widehat\alpha_p(x)),
                  \widehat\theta_p(x)\bigr).
\label{eq:model-features}
\end{equation}

The crossover surprise asks whether recombining two sampled orders yields an
order of predictable cost, which is the step GEQO relies on. The probe draws
$P=2n$ pairs of parents from the sampled orders and produces one child per
pair with edge-recombination crossover (ERX), which builds a child order from
adjacencies of both parents and is the recombination operator of PostgreSQL's
GEQO~\cite{postgresql2025geqo}. The surprise of child $j$ and the crossover
surprise are
\begin{equation}
S_j=f(\pi'_j)-\frac{f(\pi_{a_j})+f(\pi_{b_j})}{2},\qquad
\widehat\theta_p=\frac{2s_S^2}{s_f^2},
\label{eq:probe-theta}
\end{equation}
where $s_f^2$ is the sample variance of the scores of the sampled orders and
$s_S^2$ the sample variance of the surprises. As a variance ratio, the
crossover surprise ranges from zero upward and can exceed one. Its error
decreases as $O(1/\sqrt{P})$, so $P=2n$ gives an error of order $1/\sqrt n$ at
a probe cost that is linear in $n$, and $R$ is sized to supply about $P$
orders for the parent pool and the variance $s_f^2$.

The three features relate to different solvers. The relation
count sets the size of the search space and thus the planning time of DP. The
tail spacing indicates how close the best sampled orders lie to the minimum.
A small value means that many sampled orders lie near the best observed cost,
so random sampling has many chances to find a competitive order, whereas a
large value means that such orders are rare. The crossover surprise indicates
whether recombination preserves useful structure. A small value means that
children cost about as much as their parents on average, as expected when the
orders share reusable adjacencies, whereas a large value means that join
decisions interact, so that good parents can produce poor children.

JOB queries 28a and 33c both join 14 relations,
but 28a has a crossover surprise of 3.55, about twice the 1.79 of 33c. GEQO
needs 18.1\,s on 28a, where DP needs 10.2\,s, but only 0.99\,s on 33c, where DP
needs 1.34\,s. GEQO is thus the slowest solver on the query with the higher
crossover surprise, a difference that the identical relation counts cannot
reveal.

\subsection{Clusters and Dominance}
\label{sec:clusters}

Assigning one solver to a group of queries is safe only if that solver wins
on the group as a whole and the group does not consist of variants of a
single template. The first offline layer therefore clusters the training
queries of each workload $b$ and accepts a cluster winner only under both
conditions. It uses the training queries with complete probe features and a
configured subset of the model vector $z_p$, which in the configuration of
Section~\ref{sec:primary-results} consists of $n$ and
$\log(1+\widehat\alpha_p)$. Coordinates with zero variance in training are
dropped, and the remaining ones are scaled either by the training mean and
standard deviation or, for robust scaling, by the median and the
interquartile range. $K$-means with squared Euclidean distance, a fixed seed,
and 20 initializations then assigns each training query $i$ to a cluster
$c_i$.

For cluster $k$ and solver $m$, let $C_k(m)=\sum_{i:c_i=k}T_{im}$ denote the
total end-to-end time of solver $m$ in the cluster, let $m_k$ be the solver
with the lowest total, and let $C_k^{(2)}$ be the second lowest total. The
cluster is dominant when
\begin{equation}
1-C_k(m_k)/C_k^{(2)}\ge\delta
\quad\text{and}\quad
\lvert\text{template groups in }k\rvert\ge s_b,
\label{eq:dominance}
\end{equation}
and mixed otherwise. The margin $\delta$ is a hyperparameter, and the support
$s_b$ counts distinct template groups (Section~\ref{sec:protocol}), with a
threshold of five for JOB and JOB-Complex and three for IMDb-CEB. Exact ties
are resolved in the order DP, GEQO, MCTS. A dominant cluster stores $m_k$ as its cluster winner.

\subsection{Cluster Classifier}

Online, a query has to be mapped to its cluster with a few comparisons, and a
wrong mapping costs nothing when it leaves the solver unchanged but can cost
orders of magnitude when it does not. The second offline layer is therefore a
shallow tree that predicts the cluster from the scaled probe features and is
trained on a loss in end-to-end time rather than on cluster labels. The action of a cluster is its winner if the cluster is dominant and
$\pi_{\PG}$ otherwise. For training query $i$ and candidate cluster $k$, the
loss is the increase in end-to-end time when the query receives the action of
cluster $k$ instead of the action of its own cluster $c_i$. A decrease counts
as zero, so this positive-excess loss ignores gains from a wrong prediction.
Each leaf receives the cluster with the lowest summed loss. The tree tests at
most eight rank-spaced split points per feature and keeps at least
$\max(2,\lceil0.05N\rceil)$ of the $N$ training queries in every leaf.
Remaining ties are resolved first by the number of cluster errors and then by
the cluster ID.

\subsection{Decision for an Incoming Query}

The router acts once per query, after the optimizer has estimated the
cardinalities of the base relations and before join enumeration starts, so
the chosen solver plans the query as usual. It computes the probe features of
an incoming query $x$ from the optimizer's cardinality estimates and the join
graph, applies the stored scaling, and evaluates the tree, which returns a
cluster $\widehat c_b(x)$. The router then chooses the solver
\begin{equation}
\pi_b(x)=
\begin{cases}
m_{\widehat c_b(x)}, & \widehat c_b(x)\text{ is dominant},\\
\pi_{\PG}(x), & \text{otherwise},
\end{cases}
\label{eq:policy}
\end{equation}
where $m_c$ is the winner of cluster $c$ from
Equation~\eqref{eq:dominance}, and a query without probe features also
receives $\pi_{\PG}(x)$. The model stored for a workload consists of the
scaling parameters, the tree, and the winner or mixed label of every cluster.
A deployment configures one such model per workload, and without the router
PostgreSQL behaves exactly as $\pi_{\PG}$.

\section{Experimental Methodology}
\label{sec:method}

\subsection{Hardware and Workloads}

All measurements ran on a compute node with an AMD EPYC 9654 (Zen~4)
processor providing 96 vCPUs at 2.4\,GHz and 192\,GB of RAM, running Ubuntu
24.04 LTS and PostgreSQL built from its master branch (version 19devel,
February 2026).

The workloads combine a standard benchmark, a set of hard queries, and a
generated workload with many variants per template. JOB is the 113-query Join
Order Benchmark over IMDb~\cite{leis2015good}, and JOB-Complex contains queries with more complex
structure~\cite{wehrstein2150job}, of which one of 30 is excluded because both
MCTS and PostgreSQL with GEQO exceeded the three-minute timeout on it.
IMDb-CEB comes from the IMDb workload of the Cardinality Estimation
Benchmark~\cite{negi2021flowloss}, and the evaluation uses 200 of its queries
with at least nine relations, sampled in nearly equal numbers from 14
templates. The three workloads, whose sizes are listed in Table~\ref{tab:data}, are
separate query distributions, each with its own model.

\begin{table}[H]
\centering
\caption{Workloads in the evaluation. Groups are template groups, and solver
runs count three solvers and three repetitions per query.}
\label{tab:data}
\small
\begin{tabular*}{\textwidth}{@{\extracolsep{\fill}}lrrr@{}}
\toprule
Workload & Queries & Groups & Solver runs \\
\midrule
JOB & 113 & 33 & 1,017 \\
JOB-Complex & 29 & 29 & 261 \\
IMDb-CEB & 200 & 14 & 1,800 \\
\midrule
Total & 342 & 76 & 3,078 \\
\bottomrule
\end{tabular*}
\end{table}

To reduce timing noise, the end-to-end time of each query and solver is the
median of three measurements. The MCTS solver uses the negative-log reward, a tree depth of
10 before a rollout is forced, 10 best-plan replay candidates per restart
phase, and 480 iterations in the first of its Luby restart
phases~\cite{burlakov2026practical}. Each query record also contains $n$,
$\widehat\alpha_p$, $\widehat\theta_p$, and the probe time, the latter taken
from runs separate from the solver runs.

\subsection{Repeated Random Splits}
\label{sec:protocol}

Because every solver has been measured on every query, policies can be
compared by replaying recorded end-to-end times instead of running them
again. The evaluation does so over 50 random splits per workload, with split
seeds from 1 to 50. Each split holds out a random 20\% of
the queries, fits the scaling, the clusters, the cluster winners, the
dominance labels, and the tree on the remaining 80\%, and then routes the
held-out queries. Template groups enter only the support count of
the dominance test. A template group is the set of letter variants
of a JOB template, such as 28a, 28b, and 28c, the set of hashed variants of an
IMDb-CEB template, or a single JOB-Complex query.

The router configuration is chosen once on JOB by a grid search over 20 random splits (Section~\ref{sec:primary-results}) and applied unchanged to all three workloads.

Simple policies show how much of the gain a choice without probe features
already achieves, and they run on the same 50 splits
(Table~\ref{tab:controls}). Always-DP, always-GEQO, and always-MCTS run a single solver on every query. A
learned threshold runs DP below a relation count $\tau$ and another solver at
$\tau$ or above, with $\tau$ chosen on the training queries of each split. The
learned DP/MCTS threshold
\begin{equation}
\pi_{\tau}(x)=
\begin{cases}
\DP, & n(x)<\tau,\\
\MCTS, & n(x)\ge \tau
\end{cases}
\label{eq:threshold}
\end{equation}
chooses $\tau$ from 8 to 12, and the learned DP/GEQO threshold chooses it from
0 to 20. Two oracles use the held-out times, the per-count oracle choosing the
fastest solver for each relation count and the per-query oracle the fastest
solver for each query. They bound what a choice by relation count and a choice per query can achieve.

\subsection{Metrics}
\label{sec:metrics}

The metrics charge the router for everything it adds to a query. The router
time of a query is its end-to-end time with the chosen solver plus
its probe time if the query passed through the tree, while the default rule
incurs no probe time. Classifier time was not measured and is set to zero.
The held-out time of a policy is the total router time, or end-to-end time
for the baselines, over the held-out queries of a split, and its reduction is
the fraction by which it falls below the held-out time of the default rule,
so a negative reduction is an increase. Reported reductions are means over the 50 splits. For the comparison with the default rule (Table~\ref{tab:primary}), three
more quantities are averaged over the splits. They are the 90th and 95th
percentiles of the per-query reduction (P90 and P95), the near-oracle share
of held-out queries whose time is within 5\% of the oracle time
$T_i^*=\min_m T_{im}$, and the number of held-out queries that fall back to
the default rule. For the feature ablation, the spread of the reduction over the 50 splits is
summarized by its minimum, 5th
and 95th percentiles (P5 and P95), and maximum.

\section{Results}
\label{sec:results}

\subsection{Room for Choosing the Solver}

\begin{table*}[t]
\caption{Total end-to-end time per workload in seconds. Winners counts the
queries on which DP, GEQO, and MCTS are fastest.}
\label{tab:baselines}
\centering
\small
\begin{tabular*}{\textwidth}{@{\extracolsep{\fill}}lrrrrrr@{}}
\toprule
Workload & Winners & DP (s) & GEQO (s) & MCTS (s) & Default (s) & Oracle (s) \\
\midrule
JOB & 38/29/46 & 4041.00 & 3944.92 & \textbf{3534.52} & 4061.79 & 3311.15 \\
JOB-Complex & 7/9/13 & 2053.51 & 1935.56 & \textbf{1787.24} & 1874.02 & 1600.68 \\
IMDb-CEB & 52/57/91 & 8378.48 & 7632.37 & \textbf{7030.49} & 8009.86 & 6463.76 \\
\bottomrule
\end{tabular*}
\end{table*}

A router can gain only if the solvers win on different queries and a choice
per query beats the best single solver, and both hold on all three workloads
(Table~\ref{tab:baselines}). Every solver wins some queries of every workload, and no
single solver comes close to the per-query oracle. Always-MCTS is the best single policy on all three workloads and saves
between 5\% and 13\% of the default total, yet the per-query oracle is
another 6\% to 10\% faster. In the native portfolio the oracle likewise stays
five to six percentage points below the better of DP and GEQO, so choosing the
solver per query has room in both portfolios.

\begin{figure*}[t]
\centering
\includegraphics[width=\textwidth,trim=2.6pt 6.8pt 1.9pt 2.5pt,clip]{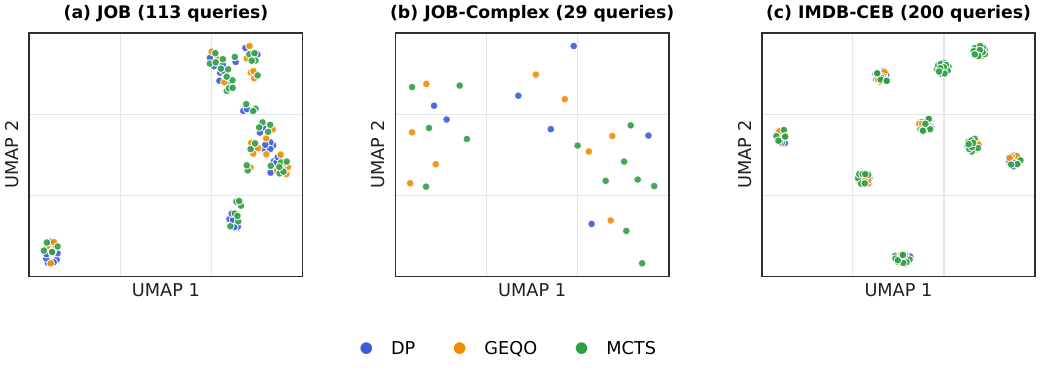}
\caption{UMAP projection of the probe features $(n,\widehat\alpha_p,\widehat\theta_p)$
of each workload to two dimensions, with colors showing the fastest solver of
each query.}
\label{fig:clusters}
\end{figure*}

For the probe features to support such a choice, queries with the same
fastest solver should lie close together in the feature space.
In Figure~\ref{fig:clusters}, the probe features of each workload are
projected to two dimensions with UMAP, and each query is colored by its
fastest solver. The queries
of JOB and IMDb-CEB form compact groups, whereas the 29 JOB-Complex queries
spread out. Within a group the fastest solver often varies, which is why the
router assigns a solver only to clusters whose winner leads in total
end-to-end time and leaves the other clusters to the default rule.

\subsection{Router Against the Default Rule}
\label{sec:primary-results}

All results use a single router configuration, found by a grid search on JOB
over 20 random 80/20 splits. The grid covers $K$ from 3 to 5 clusters, standard and robust
scaling, tree depths 2 and 3, dominance margins $\delta$ of 0, 0.05, and 0.1,
and the two feature sets $\{n,\widehat\alpha_p\}$ and
$\{n,\widehat\alpha_p,\widehat\theta_p\}$. The configuration with the highest
mean reduction on the held-out queries, which is applied unchanged to all
three workloads, has $K=4$, robust scaling, the features $n$ and
$\widehat\alpha_p$, the margin $\delta=0$, and a tree of depth 2.

\begin{table*}[t]
\caption{Router against PostgreSQL's default rule over 50 random 80/20
splits, with held-out times in seconds and metrics as defined in
Section~\ref{sec:metrics}. Near oracle is given first for the default rule
and then for the router.}
\label{tab:primary}
\centering
\small
\begin{tabular*}{\textwidth}{@{\extracolsep{\fill}}lrrrrrrr@{}}
\toprule
Workload & Default (s) & Router (s) & Mean & P90 & P95 & Near oracle & Fallback \\
\midrule
JOB & 837.18 & 737.51 & \textbf{12.26\%} & 27.89\% & 44.73\% & 59.48\%$\to$66.43\% & 1 of 23 \\
JOB-Complex & 423.00 & 426.32 & $-2.81$\% & 18.21\% & 23.70\% & 63.00\%$\to$63.67\% & 0 of 6 \\
IMDb-CEB & 1659.23 & 1511.31 & \textbf{8.53\%} & 27.88\% & 43.09\% & 56.55\%$\to$57.40\% & 2 of 40 \\
\bottomrule
\end{tabular*}
\end{table*}

Against the default rule, the router lowers mean held-out time on JOB and
IMDb-CEB (Table~\ref{tab:primary}). On JOB-Complex the mean held-out time rises although the upper
percentiles of the per-query reduction are positive, so a few expensive
regressions outweigh many smaller gains. The share of queries within 5\% of the oracle
grows by about seven percentage points on JOB and by less than one point
elsewhere, and only a few held-out queries per split fall back to the default
rule. All router times include the probe, which takes a median of about
20\,$\mu$s and at most 57\,$\mu$s per query, or at most 5\,ms per workload.

\begin{table}[t]
\centering
\caption{Mean reduction of held-out time relative to PostgreSQL's default rule
on the same 50 splits. Negative values are increases.}
\label{tab:controls}
\small
\begin{tabular*}{\textwidth}{@{\extracolsep{\fill}}lrrr@{}}
\toprule
Policy & JOB & JOB-Complex & IMDb-CEB \\
\midrule
Default rule & 0.00\% & 0.00\% & 0.00\% \\
\midrule
\multicolumn{4}{@{}l}{\emph{DP, GEQO, and MCTS}} \\
Always-MCTS & 12.12\% & $-1.06$\% & 11.33\% \\
Learned DP/MCTS threshold & 10.83\% & 0.91\% & 11.98\% \\
Router & 12.26\% & $-2.81$\% & 8.53\% \\
\midrule
\multicolumn{4}{@{}l}{\emph{DP and GEQO only}} \\
Always-DP & 0.59\% & $-15.11$\% & $-5.70$\% \\
Always-GEQO & 4.15\% & $-2.67$\% & 4.86\% \\
Learned DP/GEQO threshold & 3.93\% & $-0.76$\% & 4.20\% \\
Router & 3.02\% & $-7.72$\% & 3.94\% \\
\midrule
\multicolumn{4}{@{}l}{\emph{Oracles, DP, GEQO, and MCTS}} \\
Per-count oracle & 17.10\% & 11.51\% & 15.17\% \\
Per-query oracle & 18.61\% & 12.35\% & 18.97\% \\
\bottomrule
\end{tabular*}
\end{table}

Part of this gain could also come from a better fixed policy, such as
running MCTS everywhere. The router is therefore compared with the baseline policies on the same
splits (Table~\ref{tab:controls}). On JOB, the router reaches the gain of always-MCTS, the best single solver there,
without being told in advance which solver is best, since its cluster winners
come from the training queries. On IMDb-CEB, the learned DP/MCTS threshold and
always-MCTS lead the router by about three percentage points, and with DP and
GEQO only, always-GEQO leads it on all three workloads. On JOB-Complex, only
the learned DP/MCTS threshold avoids a regression. No evaluated policy is best on all three workloads, so even a fixed policy
has to be chosen per workload, and the router makes this choice from the
training queries. The per-query oracle reaches about 19\% on JOB and IMDb-CEB, above every
evaluated policy.

\subsection{Contribution of Each Feature}
\label{sec:feature-results}

\begin{table*}[t]
\caption{IMDb-CEB feature ablation with DP, GEQO, and MCTS. Reduction of
held-out time relative to the default rule over 50 splits
(Section~\ref{sec:metrics}).}
\label{tab:ceb-ablation-mcts}
\centering
\small
\begin{tabular*}{\textwidth}{@{\extracolsep{\fill}}lrrrrrr@{}}
\toprule
Features & Router (s) & Mean & Min & P5 & P95 & Max \\
\midrule
Default rule & 1659.23 & 0.00\% & 0.00\% & 0.00\% & 0.00\% & 0.00\% \\
$\widehat\alpha_p$ & 1476.19 & 10.74\% & $-3.38$\% & 1.38\% & 19.17\% & 20.80\% \\
$\widehat\theta_p$ & 1472.45 & 10.95\% & $-3.38$\% & 2.79\% & 18.23\% & 21.24\% \\
$\widehat\alpha_p+\widehat\theta_p$ & 1454.73 & 12.01\% & $-3.38$\% & 1.23\% & \textbf{21.59\%} & \textbf{23.61\%} \\
$n$ & 1542.72 & 6.76\% & $-3.38$\% & $-1.85$\% & 14.44\% & 16.26\% \\
$n+\widehat\alpha_p$ & 1511.31 & 8.53\% & $-3.30$\% & $-0.73$\% & 19.80\% & 22.17\% \\
$n+\widehat\theta_p$ & 1481.28 & 10.55\% & \textbf{0.49\%} & 3.23\% & 18.93\% & 23.20\% \\
$n+\widehat\alpha_p+\widehat\theta_p$ & \textbf{1451.22} & \textbf{12.24\%} & $-3.38$\% & \textbf{3.29\%} & 21.28\% & 23.27\% \\
\bottomrule
\end{tabular*}

\vspace{0.8\baselineskip}

\caption{IMDb-CEB feature ablation with DP and GEQO only, with columns as in
Table~\ref{tab:ceb-ablation-mcts}.}
\label{tab:ceb-ablation-no-mcts}
\begin{tabular*}{\textwidth}{@{\extracolsep{\fill}}lrrrrrr@{}}
\toprule
Features & Router (s) & Mean & Min & P5 & P95 & Max \\
\midrule
Default rule & 1659.23 & 0.00\% & 0.00\% & 0.00\% & 0.00\% & 0.00\% \\
$\widehat\alpha_p$ & 1608.50 & 2.86\% & $-7.29$\% & $-4.65$\% & 9.69\% & 15.18\% \\
$\widehat\theta_p$ & \textbf{1589.08} & \textbf{4.08\%} & \textbf{$-5.45$\%} & \textbf{$-2.54$\%} & \textbf{14.03\%} & 16.73\% \\
$\widehat\alpha_p+\widehat\theta_p$ & 1613.27 & 2.54\% & $-7.29$\% & $-3.88$\% & 8.82\% & 15.18\% \\
$n$ & 1594.59 & 3.80\% & $-13.10$\% & $-2.87$\% & 12.86\% & 15.18\% \\
$n+\widehat\alpha_p$ & 1591.17 & 3.94\% & $-6.11$\% & $-3.31$\% & 12.83\% & 15.18\% \\
$n+\widehat\theta_p$ & 1594.90 & 3.81\% & $-13.10$\% & $-2.87$\% & 11.47\% & 15.18\% \\
$n+\widehat\alpha_p+\widehat\theta_p$ & 1601.83 & 3.20\% & $-7.67$\% & $-4.25$\% & 11.67\% & \textbf{18.07\%} \\
\bottomrule
\end{tabular*}
\end{table*}

To measure how much each probe feature contributes, the router configuration
is kept fixed and only the feature set is varied on IMDb-CEB
(Tables~\ref{tab:ceb-ablation-mcts} and~\ref{tab:ceb-ablation-no-mcts}).
With MCTS in the portfolio, every feature set lowers mean held-out time, and
either probe feature alone already outperforms the relation count alone, at
about 11\% against 7\%. The full set of three features gives the largest
mean reduction, 12.2\%. The probe features thus carry information about IMDb-CEB queries that the
relation count lacks, and combining the tail spacing with the crossover
surprise gives more than either alone. The configuration chosen on JOB
uses only the relation count and the tail spacing and reaches 8.5\% here, so
choosing the feature set per workload could raise the gain of the router
further. The relation count alone and the relation count with the tail
spacing still raise held-out time on at least 5\% of the splits.

Without MCTS, the router chooses only between DP and GEQO, and the mean
reductions shrink to between 2.5\% and 4.1\%. The crossover surprise alone does best in this portfolio, as expected for a
feature that measures how well the recombination step of GEQO works on a
query (Section~\ref{sec:probe}). Every feature set raises held-out time on at
least 5\% of the splits, so with the native portfolio the gain of the router
is smaller and varies more between splits than with MCTS.

\subsection{Reproducibility}
\label{sec:reproducibility}

The router code, the recorded measurements, the grid search of
Section~\ref{sec:primary-results}, and the scripts that produce
Tables~\ref{tab:data}, \ref{tab:baselines}, \ref{tab:primary},
\ref{tab:ceb-ablation-mcts}, and~\ref{tab:ceb-ablation-no-mcts} are available
at \url{https://anonymous.4open.science/r/qorouter-anon}.

\section{Limitations and Future Work}
\label{sec:limitations}

The evaluated workloads are small, with 33 template groups in JOB, 29 queries
in JOB-Complex, and 14 templates in IMDb-CEB, which limits how rich a model
they can support without overfitting. JOB-Complex consists of hand-crafted hard queries, so its results describe
worst-case solver behavior more than the behavior of a learned router. The
replay combines probe times and solver times from separate runs, and
classifier time was not measured.

A first direction for future work is the selection of training data. The
evaluation labels every query with all three
solvers and three repetitions, which suits replay but exceeds what training
needs. In a replay with labels for only half of the training template
groups, the router still lowers mean held-out time by 9.0\% on JOB and 8.0\%
on IMDb-CEB, compared with 12.3\% and 8.5\% with all labels. A deliberate
selection would measure
alternative solvers only where the router lacks evidence, for example for
templates whose cluster is not yet dominant, and would reuse the end-to-end
times that the default rule already produces in normal operation.

Another direction is the handling of mixed clusters. The router currently
leaves their queries to the default rule, which ensures that every
difference from PostgreSQL comes from the cluster winners but also caps the
gain. On IMDb-CEB the learned DP/MCTS threshold outperforms the router by
about three percentage points (Table~\ref{tab:controls}), so a learned
threshold, a finer clustering, or a per-query choice inside mixed clusters
could recover part of this gap.

\section{Conclusion}
\label{sec:conclusion}

Join-order solvers differ in how they respond to the distribution of plan
costs over join orders, and the relation count that PostgreSQL routes by does
not describe this distribution. Two probe features computed in microseconds
describe it well enough for a router that acts only on clusters with a clear
winner to lower mean end-to-end time on JOB and IMDb-CEB. The per-query oracle is still about six percentage points ahead of the router
on JOB and ten on IMDb-CEB, and the first places to close this gap are the
queries of mixed clusters and the choice of training data.

\bibliographystyle{unsrtnat}
\bibliography{refs}

\end{document}